\documentclass[fleqn,10pt]{wlscirep}
\usepackage[utf8]{inputenc}
\usepackage[T1]{fontenc}
\usepackage{url}
\usepackage{color}

\title{Accelerating discovery of infrared nonlinear optical materials with large shift current via high-throughput screening}

\author[1]{Aiqin Yang}
\author[2,3]{Dian Jin}
\author[2,4]{Mingkang Liu}
\author[2,*]{Daye Zheng}
\author[5,6,*]{Qi Wang}
\author[2,5,7,8,*]{Qiangqiang Gu}
\author[1,5,6,9,*]{Jian-Hua Jiang}
\affil[1]{School of Physical Science and Technology, Soochow University, Suzhou 215006, China}
\affil[2]{AI for Science Institute, Beijing 100080, China}
\affil[3]{Research Institute for Intelligent Wearable Systems, The Hong Kong Polytechnic University, Hong Kong SAR, China}
\affil[4]{Department of Mechanical Engineering, National University of Singapore, Singapore 117575, Singapore}
\affil[5]{Suzhou Institute for Advanced Research, University of Science and Technology of China, Suzhou 215123, China}
\affil[6]{School of Biomedical Engineering, Division of Life Sciences and Medicine, University of Science and Technology of China, Hefei 230026, China}
\affil[7]{School of Artificial Intelligence and Data Science, University of Science and Technology of China, Hefei 230026, China}
\affil[8]{Suzhou Big Data \& AI Research and Engineering Center, Suzhou 215123, China}
\affil[9]{School of Physical Sciences, University of Science and Technology of China, Hefei 230026,China}

\affil[*]{Corresponding author. Email: zhengdy@aisi.ac.cn (Daye Zheng)}
\affil[*]{Corresponding author. Email: qiw@ustc.edu.cn (Qi Wang)}
\affil[*]{Corresponding author. Email: guqq@ustc.edu.cn (Qiangqiang Gu)}
\affil[*]{Principal Corresponding author. Email: jhjiang3@ustc.edu.cn (Jian-Hua Jiang)}

\begin{abstract}

Discovering nonlinear optical (NLO) materials with strong shift current response, particularly in the infrared (IR) regime, is essential for next-generation optoelectronics yet remains highly challenging in both experiments and theory, which still largely relies on case by case studies. Here, we employ a high-throughput screening strategy, applying a multi-step filter to the Materials Project database (>154,000 materials), which yielded 2,519 candidate materials for detailed first-principles evaluation. From these calculations, we identify 32 NLO materials with strong shift current response ($\sigma$ > 100 $\mu A/V^2$). Our work reveals that layered structures with $C_{3v}$ symmetry and heavy $p$-block elements (e.g. Te, Sb) exhibit apparent superiority in enhancing shift current. More importantly, 9 of these compounds show shift current response peaks in the IR region, with the strongest reaching 616 $\mu A/V^2$, holding significant application potential in fields such as IR photodetection, sensing, and energy harvesting. Beyond identifying promising candidates, this work establishes a comprehensive and high-quality first-principles dataset for NLO response, providing a solid foundation for future AI-driven screening and accelerated discovery of high-performance NLO materials, as demonstrated by a prototype machine-learning application.
\end{abstract}
\begin{document}

\flushbottom
\maketitle

\thispagestyle{empty}

\section*{Introduction}

\begin{figure}[!ht]
    \centering
    \includegraphics[width=0.8\textwidth]{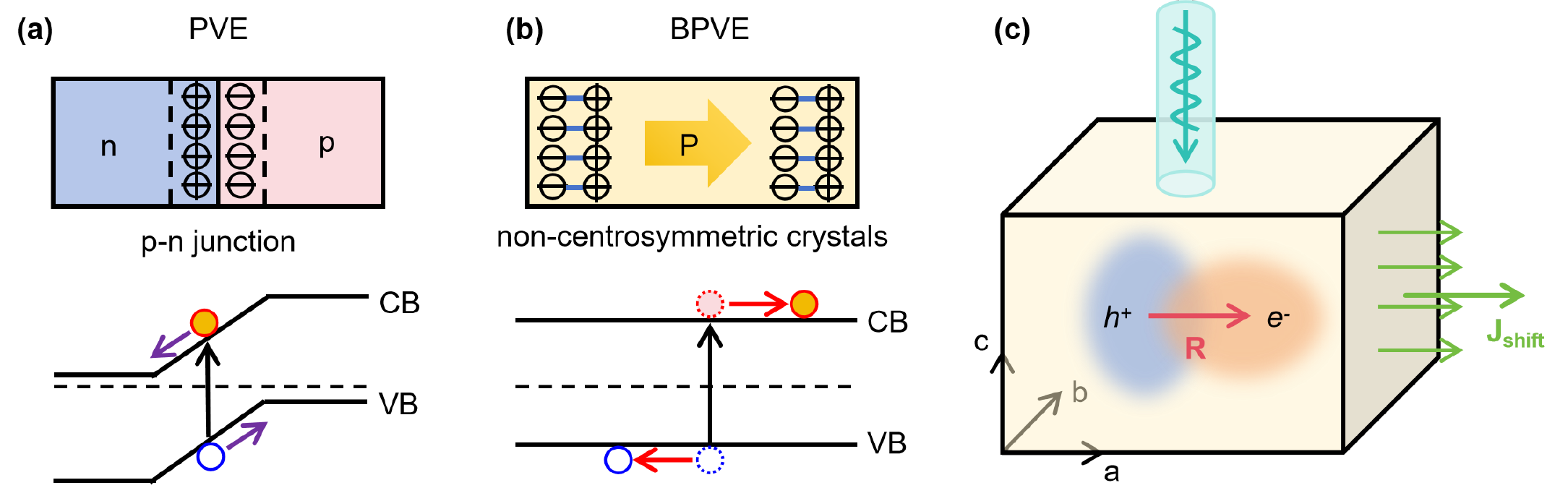}
    \caption{Schematic illustration of traditional photovoltaic effect (PVE) and bulk photovoltaic effect (BPVE). (a) Traditional PVE devices, such as p-n junctions, utilize the built-in electric field near the junction interface to separate photo-excited carriers. (b) BPVE is an intrinsic photovoltaic phenomenon occurring in non-centrosymmetric materials, relying on spontaneous polarization induced by structural asymmetry to separate photo-excited charge carriers without requiring heterojunctions or interfaces. (c) The shift current mechanism in real space on light-induced interband excitation. In non-centrosymmetric crystals, the displacement of the electron cloud during the excitation of electrons from the valence band to the conduction band generates a net current known as shift current.
    }
    \label{fig:bpve}
\end{figure}

The efficient conversion of light into electricity underpins a broad range of technologies, from imaging and communications to biosensing and renewable energy~\cite{Krishna_2005,Brongersma_2015}. Beyond the traditional photovoltaic effect (PVE) based on $p$–$n$ junctions (Figure~\ref{fig:bpve}(a)), whose open-circuit voltage is fundamentally bounded by the material's bandgap~\cite{1954_p-n_junction,2021_photo-voltage,SQ_limit_1961,SQ_limit_2016}, the bulk photovoltaic effect (BPVE) offers a distinct, interface-free mechanism enabled by inversion-symmetry breaking (Figure~\ref{fig:bpve}(b)). The BPVE occurring in non-centrosymmetric materials, a second-order nonlinear photoelectric response, can generate steady direct-current photocurrent and above-bandgap photovoltages~\cite{BPVE_Fridkin_2001,2022_Review_BPVE,2023_Review_BPVE}, making them a promising platform for next-generation nonlinear optoelectronics. 
Microscopically, the photocurrent observed in BPVE is understood as arising from two primary mechanisms: ballistic current and shift current~\cite{2023_Review_BPVE}. Ballistic current is a more extrinsic mechanism involving complex scattering processes~\cite{2021_ballistic_PhysRevLett.126.177403,2021_ballistic_PhysRevB.104.235203}. In contrast, shift current is widely considered an intrinsic mechanism, largely insensitive to scattering~\cite{PhysRevB.101.235448,PhysRevB.97.245143,JIN2023108844}. Under uniform illumination, electron excitation from valence to conduction bands in non-centrosymmetric materials produces an asymmetric momentum distribution accompanied by a shift in real-space positions~\cite{PhysRevLett.109.116601,Cook_2017}, resulting in a net shift current along the material's spontaneous polarization direction (Figure~\ref{fig:bpve}(c)). This work focuses specifically on shift current in BPVE.

Early demonstrations of shift current BPVE were reported in ferroelectric oxides, such as BaTiO$_3$~\cite{1956_BaTiO3_PhysRev.102.705,KOCH1975847_BaTiO3}, LiNbO$_3$~\cite{1974_LiNbO3_APL}, and Pb(ZrTi)O$_3$~\cite{BRODY1975193_PbTiO3}.
Due to their wide bandgap (2.7$-$4 eV), these materials typically exhibit photovoltaic response mainly in the ultraviolet (UV) to visible (VIS) spectrum. The development of emerging materials, such as two-dimensional materials and topological semimetals, has led to more extensive and in-depth studies of BPVE, extending its spectral response range into the IR and even terahertz (THz) regions. Osterhoudt et al.~\cite{Osterhoudt2019_TaAs} reported a giant shift current response in Weyl semimetal TaAs ($\sigma^{aac} \sim$ 154 $\mu A/V^2$ at 10.6 $\mu$m) exceeding conventional ferroelectrics by an order of magnitude, pushing the operational wavelength of BPVE deep into the mid-infrared (MIR). Wang et al.~\cite{Wang2024_Te_MIR} further demonstrated broadband coverage from UV to MIR (390 nm–3.8 $\mu$m) in tellurium, achieving an extraordinary photocurrent density of 70.4 A/cm$^2$ at 1.31 $\mu$m. These breakthroughs establish shift current BPVE as a powerful mechanism spanning VIS to THz wavelengths with unprecedented intensity.
IR wavelengths constitute a vital segment of the electromagnetic spectrum, characterized by low energy and strong penetration power, playing a significant role across multiple fields including industry, medicine, and military applications. Given that the BPVE generates steady direct-current photocurrent without external voltage and can theoretically surpass the Shockley-Queisser (S-Q) limit of traditional PVE, the IR shift current holds immense  potential in high-sensitivity IR detection without external bias, self-powered optoelectronic sensing, broad-spectrum energy harvesting, and deep biomedical regulation.
However, the practical application of BPVE remains constrained by low photocurrents and poor energy conversion efficiency. Therefore, identifying materials capable of strong shift current response, particularly within the IR spectrum, represents an urgent and critical research priority.

A comprehensive computational search offers a promising pathway for accelerating discovery of NLO materials with strong shift current response. 
The high-throughput screening approach has successfully identified high-performance candidates across diverse applications, from photovoltaics~\cite{Yu2012_PhysRevLett.108.068701_PV} and nonlinear optics~\cite{2024_SHG_HT_PhysRevMaterials.8.085202} to power electronics~\cite{Chen2025_npj_HTscreening}, thermoelectrics~\cite{Gorai2017_thermoelectric} and field-effect transistors~\cite{Li2024_nc_FETs}, far surpassing traditional trial-and-error methods.
However, current high-throughput efforts based on BPVE have been limited to either specific structural families, such as the two-dimensional CuXX'Y materials~\cite{2023_CuXX'Y_PhysRevMaterials.7.074001}, or small databases, such as the 326 non-centrosymmetric entries in the C2DB~\cite{Sauer2023_npj_efficiency_2D}. Critically, no systematic exploration of NLO shift current response has been conducted across the much broader Materials Project (MP) database, which contains more than 154,000 inorganic materials. This leaves the large structural and chemical space of MP essentially uncharted for high-performance IR shift-current materials. Meanwhile, the rapid rise of AI-assisted materials discovery highlights the indispensable role of high-quality, large-scale datasets in the development of predictive and generative models. For NLO effect, particularly the shift current, such datasets are virtually missing, causing a severe bottleneck for advancing data-driven discovery pipelines. Establishing a reliable first-principles dataset for shift current response is therefore as fundamental as improving the underlying machine-learning models themselves.

In this work, we address these gaps by performing unbiased high-throughput first-principles calculations, with the exchange-correlation functional in the Perdew-BurkeErnzerhof (PBE) form~\cite{1996_PBE}, on 2,519 non-centrosymmetric materials filtered from the MP database. From these calculations, 32 materials exhibiting strong shift current response ($\sigma >$ 100 $\mu A/V^2$) were identified. For these top 32 candidates, we further performed Heyd-Scuseria-Ernzerhof (HSE) hybrid functional~\cite{2003_HSE,Lin2020_hse} calculations to obtain more accurate bandgaps and NLO response. Ultimately, we identified 9 NLO materials that exhibit pronounced shift current peaks within the IR range as well as 18 and 5 NLO materials with peak response in the VIS and UV ranges, respectively. These IR candidate materials expand the BPVE material landscape and establish promising platforms for IR-related technologies such as IR sensing, imaging and communications.
Finally, beyond identifying promising materials, our high-throughput workflow also generates one of the most comprehensive first-principles datasets of shift-current response to date. A simple machine-learning demonstration based on DPA3-$\sigma$ model further illustrates its potential for accelerating large-scale, data-driven screening of NLO materials which can find valuable applications in lasing, photo-detection, and quantum information processing for various frequencies. Together, these efforts establish a robust foundation for advancing both physics-based and AI-driven discovery of high-performance IR shift current materials.

\section*{Results}

\subsection*{Database screening and high-throughput calculations}

\begin{figure*}[!ht]
    \centering
    \includegraphics[width=0.98\textwidth]{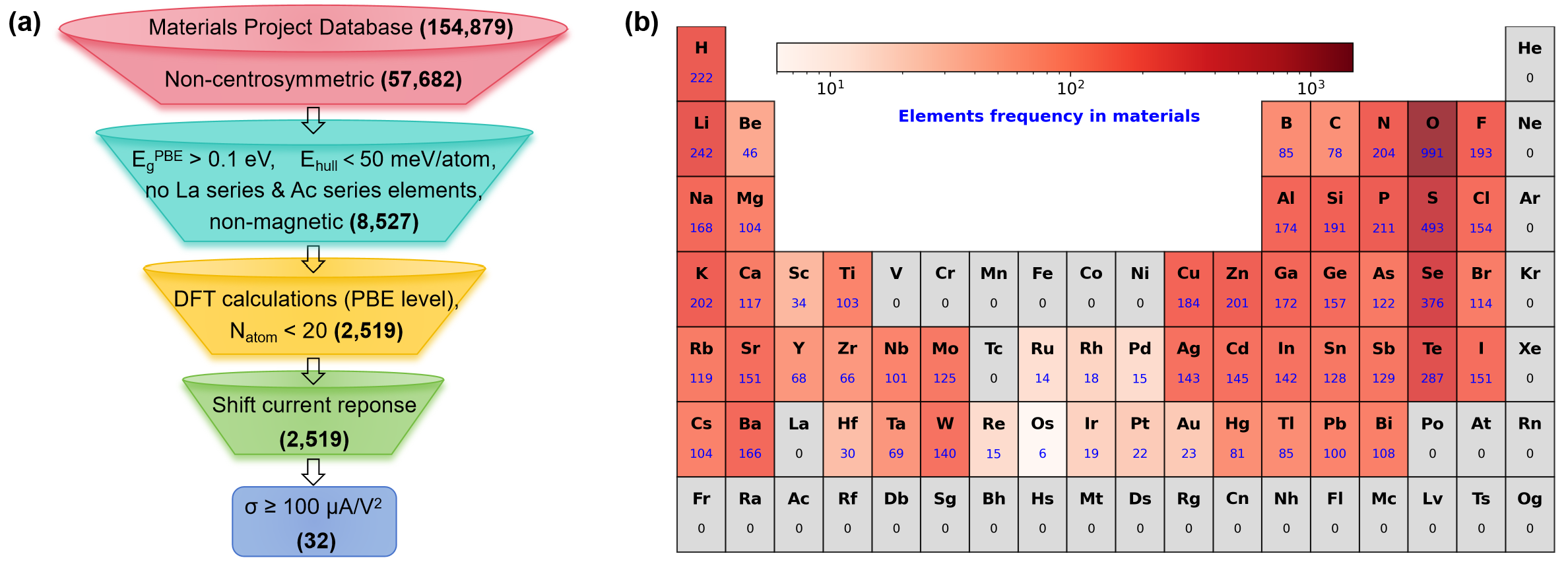}
    \caption{Database screening and high-throughput calculations. (a) The workflow of high-throughput screening process. Initially, filters based on symmetry, band gap, number of atoms, magnetic properties, and elemental composition were applied. Subsequently, DFT calculations were conducted to obtain the Hamiltonian and band structure. Following this, the DFT Hamiltonians were utilized to compute the shift current response at the PBE level. Ultimately, 32 materials with high shift current response were identified as potential candidates. Importantly, further calculations at the HSE level were performed for these target candidates to refine the results. (b) The distribution of 56 elements in the periodic table for 2,519 noncentrosymmetric materials, where the color intensity represents the frequency of occurrence in all screened materials, and the numbers below the chemical symbols show the count of occurrence of that element. Gray represents elements that did not appear in this study.}
    \label{fig: HT}
\end{figure*}

Our high-throughput screening workflow, illustrated in Figure~\ref{fig: HT}(a), commences with the 154,879 inorganic compounds in the MP database~\cite{2013_Jain_MP}.
A multi-step filtering procedure was applied to identify non-centrosymmetric, thermodynamically reasonable, and computationally tractable candidates suitable for shift current calculations. The screening criteria are as follows: (1) Symmetry: materials must be non-centrosymmetric, which yielded 57,682 structures. (2) Electronic structure: candidates were selected to have a PBE band gap E$_g$ > 0.1 eV and be non-magnetic. (3) Thermodynamic stability: structures with an energy above the convex hull $E_{hull} \le 50$ meV/atom were selected as potentially synthesizable. (4) Elemental composition: systems containing elements with partially filled 3$d$ shells (V, Cr, Mn, Fe, Co, Ni) or lanthanide/actinide elements were removed to avoid uncertainties in density functional theory (DFT) treatment of localized $d$/$f$ electrons. (5) Computational tractability: structures with fewer than 20 atoms per unit cell were retained to ensure feasible high-throughput calculations. Beyond this restriction, no further bias toward specific chemistries or structure types was imposed. 

Following these filtering steps, 2,519 candidate materials remained for first-principles evaluation. For these filtered candidates, we employed a two-stage computational strategy. First, the electronic structure and optical response for all 2,519 candidates were computed using the PBE exchange-correlation functional. Recognizing that the shift current response is highly sensitive to band structure details and that PBE tends to underestimate band gaps, we performed a second stage of calculations. Specifically, we identified the 32 most promising materials (defined in the next section as those with a PBE response $\sigma > 100 ~\mu A/V^2$) and performed more accurate HSE hybrid functional~\cite{Lin2020_hse} calculations to obtain refined electronic structures and NLO response.

\subsection*{Statistical analysis of shift current response preference}

\begin{figure*}[!ht]
    \centering
    \includegraphics[width=\textwidth]{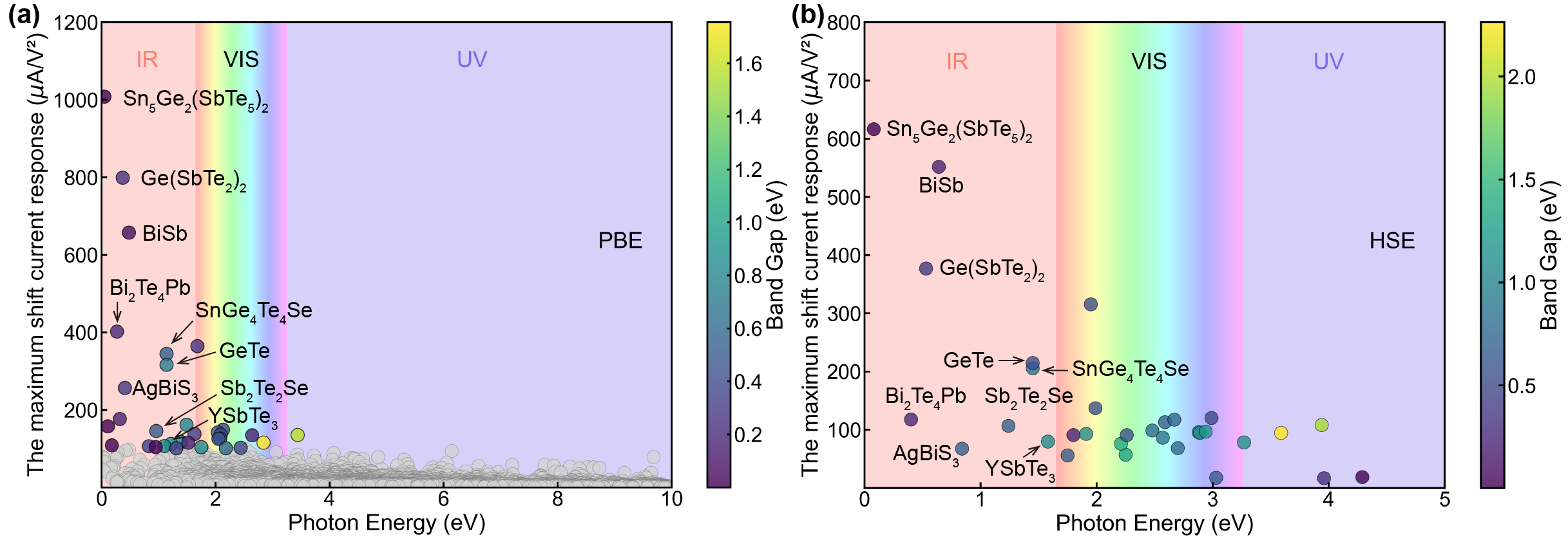}
    \caption{Maximum shift current response distribution of screened materials. The calculated maximum shift current response tensor and its corresponding photon energy at the (a) PBE and (b) HSE levels, with the colorbar revealing the band gap of 32 compounds with shift current conductivity of over 100 $\mu A/V^2$ and the gray circles indicate the remaining compounds. The background color reveals the response wavelength of the calculated materials, which falls within the infrared (IR), visible (VIS), or ultraviolet (UV) spectrum. The 9 compounds in Table~\ref{tab:sc_IR} that exhibit IR shift current response are labeled.}
    \label{fig: sc_max}
\end{figure*}

Figure~\ref{fig: HT}(b) illustrates the elemental distribution of the screened 2,519 noncentrosymmetric materials covering 56 elements across the periodic table, with color intensity representing their frequency of occurrence in the screened material and the numbers below the chemical symbols indicating their count. It can be observed that element oxygen appears most frequently (991 occurrences) among the screened materials. Except for elements excluded by the screening criteria and a few rare elements, the screened dataset spans nearly the entire periodic table, reflecting the unbiased nature of our selection process. 
The shift current response tensor $\sigma ^{abc}$ is a third-order tensor with 18 independent components. Considering the structural symmetry of crystal, $\sigma ^{abc}$ may be further simplified. Figure~\ref{fig: sc_max}(a) displays the maximum values of the shift current response tensor at the PBE level for 2,519 screened materials, along with their distributions relative to the photon energy and the band gap. Our work identified 32 candidate materials under the condition that the absolute value of the shift current tensor element exceed 100 $\mu A/V^2$. The colorbar on the right in the Figure~\ref{fig: sc_max}(a) shows the band gaps of the top 32 compounds with $|\sigma^{abc}|_\text{max} > 100$ $\mu A/V^2$, and gray circles indicate the remaining screened materials.
It can be found that the maximum shift current tensor $|\sigma^{abc}|_\text{max}$ spans widely from 0 to 1000 $\mu A/V^2$, mainly clustering below 200 $\mu A/V^2$, with corresponding photon energy ranging from 0 to 10 eV. The background color in the Figure~\ref{fig: sc_max} indicates the spectral region (IR, VIS, or UV) in which the maximum shift current response of each screened material occurs. Among these highly responsive materials, 20 compounds exhibit maximum shift current response peaks in the IR region, 10 in the VIS region, and 2 in the UV region. In contrast, previously reported materials predominantly displayed shift current response in the VIS or UV wavelength ranges. More importantly, the compounds with the strongest response include Sn$_5$Ge$_2$(SbTe$_5$)$_2$ (mp-1219067), Ge(SbTe$_2$)$_2$ (mp-1224350), BiSb (mp-1227290), Bi$_2$Te$_4$Pb (mp-1227398), and SnGe$_4$Te$_4$Se (mp-1218953), all of which exhibit shift current response in the IR region, with peak intensities exceeding 400 $\mu A/V^2$. In particular, Sn$_5$Ge$_2$(SbTe$_5$)$_2$ (mp-1219067) even presents a giant shift current response, with peak intensities reaching as high as 1000 $\mu A/V^2$ at the PBE level.

\begin{figure*}[!ht]
    \centering
    \includegraphics[width=0.98\textwidth]{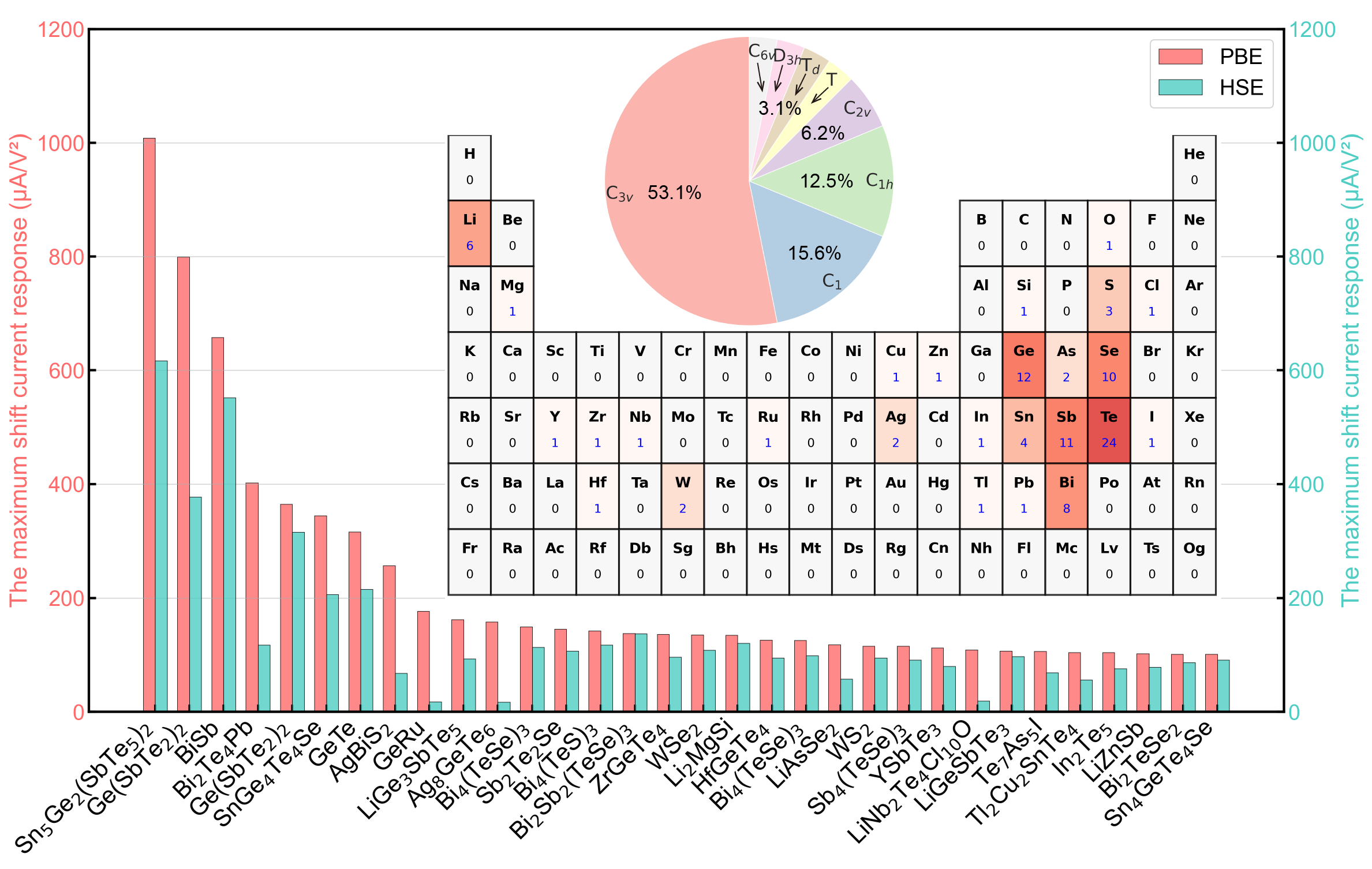}
    \caption{The calculated the maximum peak of the shift current at the PBE and HSE levels for 32 compounds with higher nonlinear optical response. The pie chart in the inset displays the relative proportions of 8 noncentrosymmetric point groups among these 32 compounds, while the periodic table on the right illustrates their elemental distribution and frequencies.}
    \label{fig: sc_comparison}
\end{figure*}

For the top 32 candidates with large shift current, there are 30 bulk materials and 2 two-dimensional materials, covering a wide range of compounds from binary to pentanary compounds. The most preferred combination of elements is Bi-Se-Te, which appears three times in 32 compounds. Next are the combinations of Ge-Sb-Te, Ge-Se-Sn-Te, Ge-Li-Sb-Te, and Sb-Se-Te, each of which appears twice in 32 compounds. Obviously, the element Te has strong superiority in the enhancement of shift current. The periodic table in the inset of Figure~\ref{fig: sc_comparison} shows the elemental distribution of these 32 compounds, and the blue numbers below the chemical symbols represent the number of occurrences of each element in different materials.  The results illustrate the superiority of Te, Sb, Ge, and Se elements in enhancing shift current response. This aligns with previous studies that have reported significant BPVE in tellurium films~\cite{PhysRevB.103.245415}. In addition, Qian et al.~\cite{Qian2023_npj} also discovered that elemental two-dimensional ferroelectrics (As, Sb, Bi) exhibit strong shift current response in the VIS range using first-principles calculations. The pie chart in the inset of Figure~\ref{fig: sc_comparison} reveals that these 32 compounds contain a total of eight point group symmetries, with the $C_{3v}$ point group accounting for 53.1\% and the $C_1$ and $C_{1h}$ point groups also accounting for a significant proportion. In particular, the top seven compounds with the largest shift current all crystallize in the trigonal $R3m$ space group (No. 160), belonging to the $C_{3v}$ symmetry point. Based on the above analysis, we can conclude that the PBE screening successfully identifies a key design principle: materials with the $C_{3v}$ point group and containing Te, Sb, Ge, and Se elements are more favorable for generating stronger shift current response.

Undoubtedly, the DFT calculation of the PBE level serves as an effective screening tool to identify promising candidate materials. With possible underestimation of bandgaps, we conducted further calculations using the more accurate HSE method. Figure~\ref{fig: sc_max}(b) displays the maximum values of the shift current response tensor at the HSE level for top 32 candidates, along with their distributions relative to the photon energy and the band gap. The histograms in Figure~\ref{fig: sc_comparison} provide a more intuitive comparison of the PBE and HSE results for the maximum shift current response across the top 32 compounds. As expected, the largest shift current response, calculated at the HSE level, shows a distinct reduction. For example, the maximum shift current conductivity decreased by about 39\% for Sn$_5$Ge$_2$(SbTe$_5$)$_2$ (mp-1219067) (from 1008 to 616 $\mu A/V^2$), by about 53\% for Ge(SbTe$_2$)$_2$ (mp-1224350) (from 799 to 377 $\mu A/V^2$), and by about 16\% for BiSb (mp-1227290) (from 657 to 552 $\mu A/V^2$). Even more pronounced is the reduction in Bi$_2$Te$_4$Pb (mp-1227398), which decreased from 402 to 118 $\mu A/V^2$ ($\sim$ 71\%), AgBiS$_2$ (mp-675977), which decreased from 257 to 68 $\mu A/V^2$ ($\sim$ 73\%), GeRu (mp-1025397), which decreased from 177 to 18 $\mu A/V^2$ ($\sim$ 90\%), and LiNb$_2$Te$_4$Cl$_{10}$O (mp-1235095), which decreased from 109 to 19 $\mu A/V^2$ ($\sim$ 82\%). The detailed comparison of results is presented in Supplementary Table 1. In addition to lower NLO response peaks, due to the correction of the band gap by the HSE hybrid functional, the photon energy corresponding to the maximum shift current response of most systems shifts toward higher energies, causing some compounds that were originally in the IR wavelength range to move to the VIS wavelength range (as shown in Figure~\ref{fig: sc_max}(b)). Specifically, among the top 32 candidate materials, 9 compounds have a shift current response in the IR region, 18 in the VIS region, and 5 in the UV region.

\subsection*{Candidates working in the IR region}

Based on high-throughput PBE and HSE calculations, we identified 9 materials exhibiting strong IR shift current response (Table~\ref{tab:sc_IR}). Sn$_5$Ge$_2$(SbTe$_5$)$_2$ (mp-1219067) is predicted to be metallic, while the remaining eight are narrow-gap semiconductors with bandgaps of 0.21–1.09 eV. With the exception of AgBiS$_2$ (mp-675977), all compounds crystallize in the layered trigonal $R3m$ structures with $C_{3v}$ symmetry and broken inversion symmetry. Further symmetry analysis reveals distinct tensor properties for these materials. $C_{3v}$ materials exhibit four independent, non-vanishing components of the shift current response: $xxy = yxx = -yyy$, $xxz = yyz$, $zxx = zyy$ and $zzz$. In contrast, AgBiS$_2$ (with $C_{1h}$ symmetry) shows ten non-vanishing components: $xxx$, $xxz$, $xyy$, $xzz$, $yxy$, $yyz$, $zxx$, $zxz$, $zyy$ and $zzz$. For all $C_{3v}$ compounds, the in-plane $xxy$ ($yxx/yyy$) component dominates the shift current response, with maximum values presented in Table~\ref{tab:sc_IR} (HSE results, with PBE values in parentheses).  In contrast, for AgBiS$_2$, PBE calculations predict a strong $xxx$ peak of 257 $\mu A/V^2$, which disappears at the HSE level. Instead, a new $xyy$ peak of approximately 68 $\mu A/V^2$ emerges near 0.84 eV, as shown in Figure~\ref{fig: sc_spectra_IR}(i). Among the 9 IR-active materials, BiSb (mp-1227290), GeTe (mp-938), and SnGe$_4$Te$_4$Se (mp-1218953) are thermodynamically stable, while the remaining six are metastable. Our screening successfully reproduces the shift current effect in BiSb and GeTe, consistent with prior work by Yang et al. on pressure and symmetry modulation in BiSb~\cite{2024_jpcc_BiSb} and Tiwari et al. on enhanced response in monolayer GeTe~\cite{Tiwari_2022_GeTe}.

\begin{table*}[ht]
\renewcommand{\arraystretch}{1.5}
\centering
\caption{The screened candidates at the HSE level exhibiting significant shift current response in the IR region. $E_{hull}$ represents the energy above the convex hull. $E_g$ represents the calculated bandgap value. $|\sigma^{abc}|_{\text{max}}$ is the peak value of the maximum component in the shift current spectra. $E_{photon}$ is the photon energy corresponding to the maximum response peak. Component reveals the maximum shift current tensor element. The values or components in parentheses are the results of PBE calculations.}
\label{tab:sc_IR}
\begin{tabular}{|l|c|c|c|c|c|c|c|}
\hline
Material\_id &
  Formula &
  Point group &
  \begin{tabular}[c]{@{}c@{}}$E_{Hull}$\\ ($meV$)\end{tabular} &
  \begin{tabular}[c]{@{}c@{}}$E_{\text{gap}}$\\ ($eV$)\end{tabular} &
  \begin{tabular}[c]{@{}c@{}}$|\sigma^{abc}|_{\text{max}}$\\ ($\mu A/V^2$)\end{tabular} &
  \begin{tabular}[c]{@{}c@{}}$E_{photon}$\\ ($eV$)\end{tabular} &
  Component \\ \hline
mp-1219067 & Sn$_5$Ge$_2$(SbTe$_5$)$_2$ & $C_{3v}$ & 29 & 0.00 (0.00) & 616 (1008) & 0.08 (0.05)  & $xxy=yxx=-yyy$  \\ \hline
mp-1227290 & BiSb                       & $C_{3v}$ & 0  & 0.21 (0.06) & 552 (657) & 0.64 (0.48)  & $xxy=yxx=-yyy$  \\ \hline
mp-1224350 & Ge(SbTe$_2$)$_2$           & $C_{3v}$ & 17 & 0.38 (0.26) & 377 (799) & 0.53 (0.37)  & $xxy=yxx=-yyy$  \\ \hline
mp-938     & GeTe                       & $C_{3v}$ & 0  & 0.99 (0.72) & 215 (316) & 1.45 (1.14)  & $xxy=yxx=-yyy$  \\ \hline
mp-1218953 & SnGe$_4$Te$_4$Se           & $C_{3v}$ & 0  & 0.85 (0.52) & 206 (345) & 1.45 (1.14)  & $xxy=yxx=-yyy$  \\ \hline
mp-1227398 & Bi$_2$Te$_4$Pb             & $C_{3v}$ & 27 & 0.25 (0.19) & 118 (402) & 0.40 (0.27)  & $xxy=yxx=-yyy$  \\ \hline
mp-1219475 & Sb$_2$Te$_2$Se             & $C_{3v}$ & 32 & 0.70 (0.48) & 107 (146) & 1.24 (0.96)  & $xxy=yxx=-yyy$  \\ \hline
mp-1215924 & YSbTe$_3$                  & $C_{3v}$ & 32 & 1.09 (0.87) & 80 (112)  & 1.58 (1.22)  & $xxy=yxx=-yyy$  \\ \hline
mp-675977  & AgBiS$_2$                  & $C_{1h}$ & 47 & 0.69 (0.27) & 68 (257)  & 0.84 (0.41)  & $xyy$ ($xxx$)          \\ \hline
\end{tabular}
\end{table*}

\begin{figure*}[!ht]
    \centering
    \includegraphics[width=0.99\textwidth]{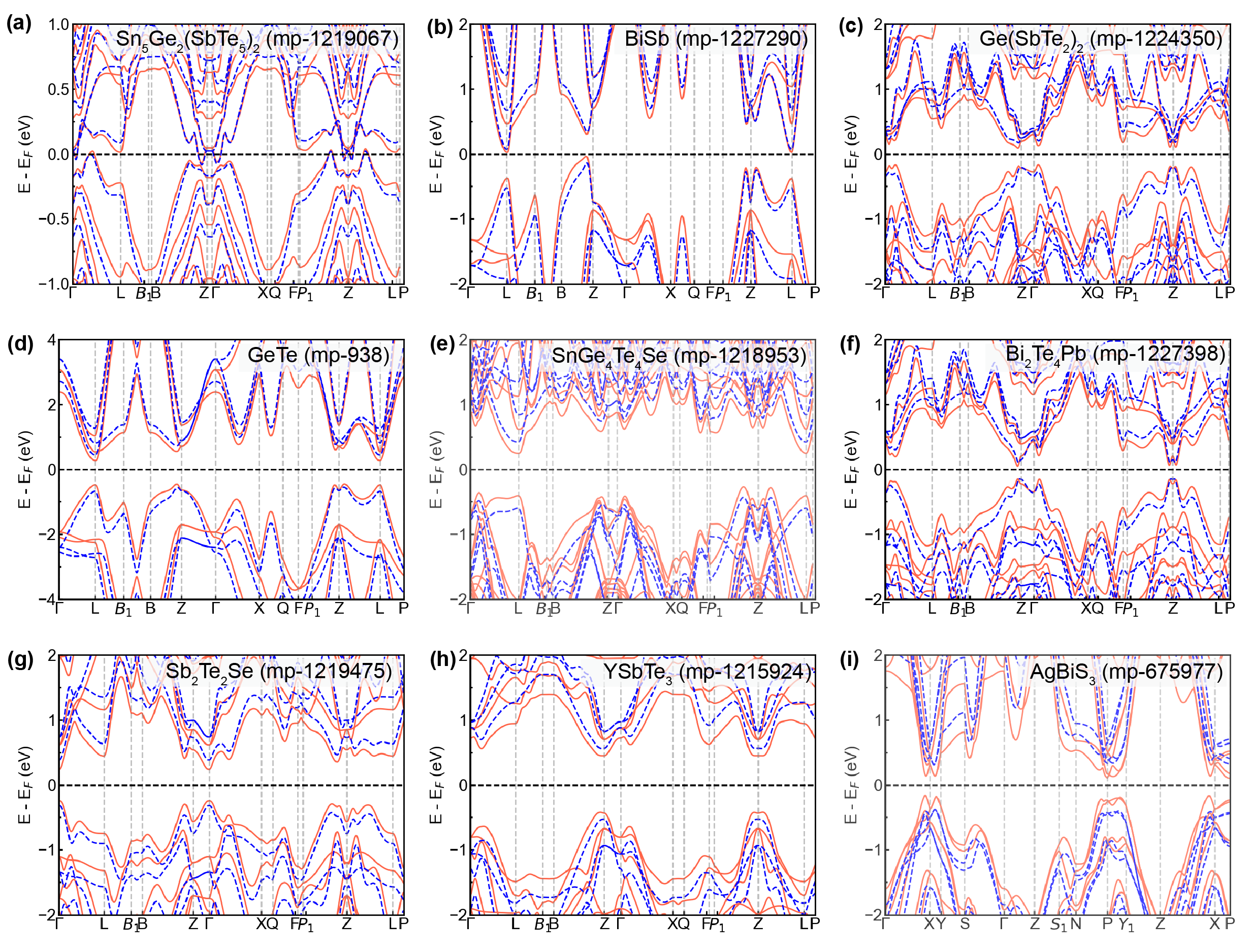}
    \caption{The band structures for the infrared candidates. (a) Sn$_5$Ge$_2$(SbTe$_5$)$_2$ (mp-1219067). (b) BiSb (mp-1227290). (c) Ge(SbTe$_2$)$_2$ (mp-1224350). (d) GeTe (mp-938). (e) SnGe$_4$Te$_4$Se (mp-1218953). (f) Bi$_2$Te$_4$Pb (mp-1227398). (g) Sb$_2$Te$_2$Se (mp-1219475). (h) YSbTe$_3$ (mp-1215924). (i) AgBiS$_2$ (mp-675977). Red solid lines and blue dashed lines correspond to PBE and HSE results, respectively.}
    \label{fig: band_IR}
\end{figure*}

\begin{figure*}[!ht]
    \centering
    \includegraphics[width=0.99\textwidth]{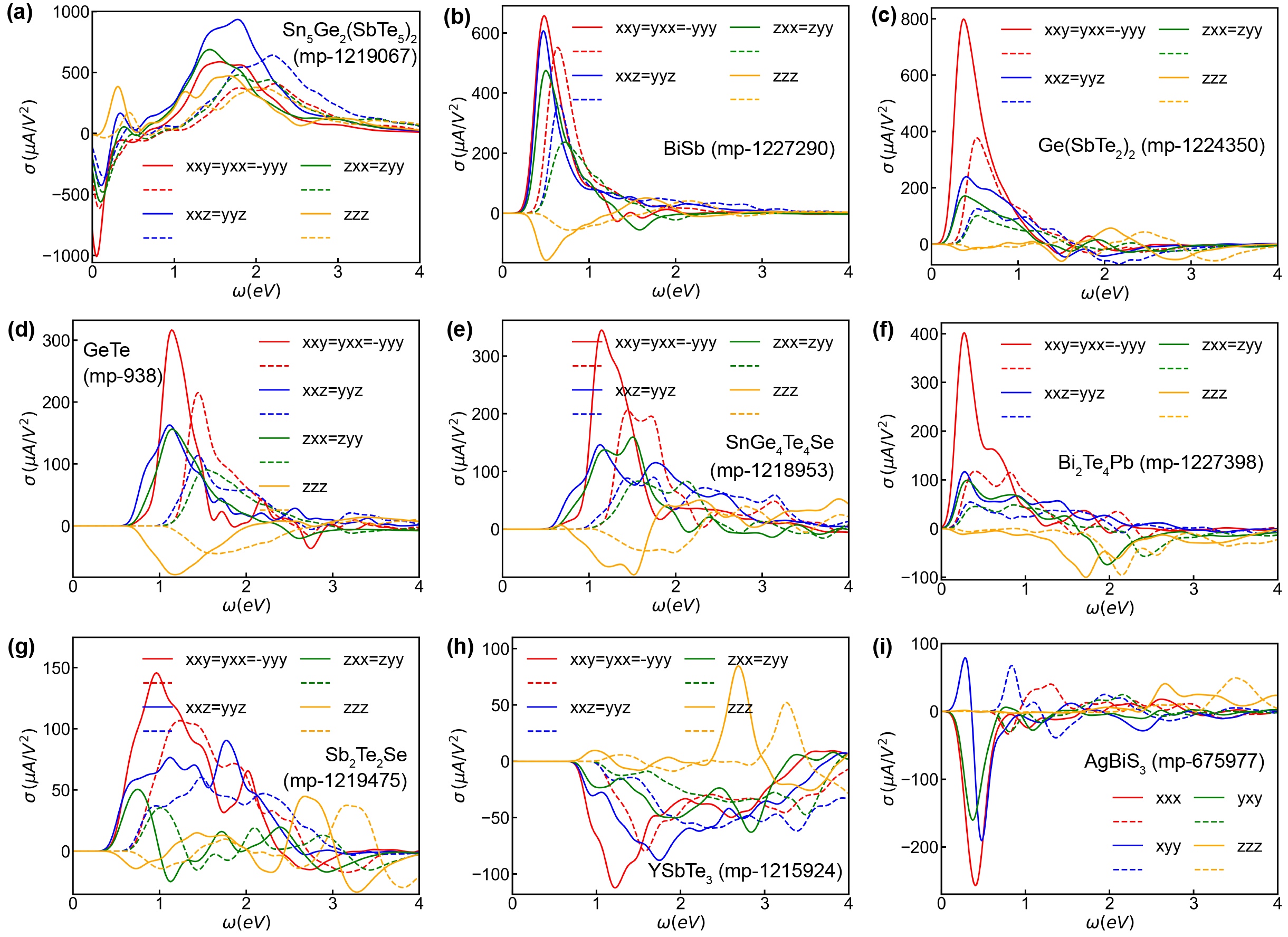}
    \caption{The shift current response spectra for the infrared candidates. For (a) Sn$_5$Ge$_2$(SbTe$_5$)$_2$ (mp-1219067), (b) BiSb (mp-1227290), (c) Ge(SbTe$_2$)$_2$ (mp-1224350), (d) GeTe (mp-938), (e) SnGe$_4$Te$_4$Se (mp-1218953), (f) Bi$_2$Te$_4$Pb (mp-1227398), (g) Sb$_2$Te$_2$Se (mp-1219475) and (h) YSbTe$_3$ (mp-1215924), red lines represent the $xxy=yxx=-yyy$ component, blue lines represent the $xxz=yyz$ component, green lines represent the $zxx=zyy$ component, orange lines represent the $zzz$ component. For (i) AgBiS$_2$ (mp-675977), red lines represent the $xxx$ component, blue lines represent the $xyy$ component, green lines represent the $yxy$ component, orange lines represent the $zzz$ component. Solid lines and dashed lines correspond to PBE and HSE results, respectively.}
    \label{fig: sc_spectra_IR}
\end{figure*}

The electronic band structures of 9 infrared NLO candidates and their corresponding shift current spectra are depicted in Figure~\ref{fig: band_IR} and Figure~\ref{fig: sc_spectra_IR}, respectively. The element-resolved and orbital-projected electronic density of states (DOS) at the HSE level are displayed in Supplementary Figure 1–3.
From the perspective of the photon energy corresponding to the maximum shift current response, four compounds exhibit MIR response, while five show near-infrared (NIR) response.

The most striking candidates is Sn$_5$Ge$_2$(SbTe$_5$)$_2$ (mp-1219067), which crystallizes in layered Te-Sb-Te-Sb-Te-Ge-Te-Ge-Te-Sn-Te-Sn-Te-Sn-Te-Sn-Te-Sn-Te sheets along the (0 0 1) direction.
Despite the MP database claiming that it has a band gap of 0.19 eV, our calculation of electronic structure  at both PBE and HSE levels reveal semimetal-like electronic characteristics, rather than a true insulating gap, as shown in Figure~\ref{fig: band_IR}(a). Specifically, no direct band crossing occurs at a single $k$ point, while small but finite gaps of $\sim$0.04 eV persist near the $\Gamma$ and Z high-symmetry points, expanding to $\sim$0.1 eV along the $\Gamma$-L and Z-L paths. 
The electronic DOS in Supplementary Figure 1(a) and Supplementary Figure 2(a) reveals that the valence band of Sn$_5$Ge$_2$(SbTe$_5$)$_2$ (mp-1219067) is dominated by Te-$p$ states, while the conduction band is primarily formed by Sb-$p$ and Ge-$p$ orbitals. This delocalized $p$-orbital character near the Fermi level is consistent with previously reported trends favoring enhanced shift current response~\cite{Tan_2016}. 
Figure~\ref{fig: sc_spectra_IR}(a) illustrates the shift current spectra at both PBE and HSE levels for Sn$_5$Ge$_2$(SbTe$_5$)$_2$ (mp-1219067). In the MIR range, Sn$_5$Ge$_2$(SbTe$_5$)$_2$ (mp-1219067) exhibits pronounced peaks across multiple tensor components: the $xxy$ ($yxx/yyy$) component reaches 616 $\mu A/V^2$ at 0.08 eV, while the $xxz$ ($yyz$) and $zxx$ ($zyy$) attain 361 $\mu A/V^2$ and 480 $\mu A/V^2$ at 0.14 eV and 0.12 eV, respectively (at the HSE level). These strong responses in the MIR range originate primarily  from robust electron transitions between the valence band and the conduction band. PBE systematically overestimates these magnitudes, as mentioned earlier.  Furthermore, Sn$_5$Ge$_2$(SbTe$_5$)$_2$ (mp-1219067) shows a strong visible-light response,  with the $xxz$ ($yyz$) component exceeding 500 $\mu A/V^2$ across the 1.7$-$2.4 eV range. 

BiSb (mp-1227290) adopts a layered hexagonal structure composed of alternating Sb and Bi atomic bilayers stacked along the $c$ axis with weak van der Waals interlayer bonding. Electronic structure calculations at both PBE and HSE levels (Figure~\ref{fig: band_IR}(b)) confirm it as an indirect-gap semiconductor with a 0.2 eV bandgap at the HSE level, consistent with previous reports~\cite{2016_PRB_BiSb,2024_jpcc_BiSb}. The valence band maximum lies along the B-Z direction, while the conduction band minimum is located near the L point, corresponding to direct gaps of 0.48 eV and 0.6 eV, respectively. 
The electronic DOS (Supplementary Figure 1(b) and Supplementary Figure 2(b)) reveals that the valence band is dominated by Sb-$p$ states with minor Bi contribution, whereas the conduction band features primarily Bi-$p$ character. This delocalized $p$-orbital contribution near the Fermi level enhances the shift current response.
The calculated shift current spectra (Figure~\ref{fig: sc_spectra_IR}(b)) reflect these electronic features: HSE corrections reduce the $\sigma^{xxz}$, $\sigma^{zxx}$, and $\sigma^{zzz}$ peaks from 609, 478, and 158 $\mu A/V^2$ (PBE) to 362, 235, and 56 $\mu A/V^2$, respectively, while blue-shifting their photon energies from 0.47, 0.50, and 0.51 eV to 0.65, 0.72, and 0.78 eV. In contrast, the in-plane $\sigma^{yyy}$ peak shows greater resilience, decreasing only modestly from 657 to 552 $\mu A/V^2$ with a blue-shift from 0.48 to 0.64 eV. This confirms that the shift current response for BiSb (mp-1227290) lies entirely within the IR spectral range.

Ge(SbTe$_2$)$_2$ (mp-1224350) and Bi$_2$Te$_4$Pb (mp-1227398) share similar seven-layer crystal structures stacked along the $z$-direction of the hexagonal unit cell in the sequence Te-Ge/Pb-Te-Sb/Bi-Te-Sb/Bi-Te sequence. Figure~\ref{fig: band_IR}(c) and \ref{fig: band_IR}(f) display respectively the band structures of Ge(SbTe$_2$)$_2$ (mp-1224350) and Bi$_2$Te$_4$Pb (mp-1227398), with band gaps of 0.38 eV and 0.25 eV, at the HSE level. The electronic DOS in Supplementary Figure 1(c, f) and Supplementary Figure 2(c, d) reveal that Te-$p$ states dominate the valence bands of both compounds, while their conduction bands are primarily formed by Sb-$p$ (Ge(SbTe$_2$)$_2$) and Bi-$p$ (Bi$_2$Te$_4$Pb) orbitals. Consequently, Bi$_2$Te$_4$Pb exhibits a lower IR shift current peak than Ge(SbTe$_2$)$_2$ (Figs.~\ref{fig: sc_spectra_IR}(c) and \ref{fig: sc_spectra_IR}(f)), likely due to its reduced band edge density of states.

GeTe (mp-938), SnGe$_4$Te$_4$Se (mp-1218953), Sb$_2$Te$_2$Se (mp-1219475), and YSbTe$_3$ (mp-1215924) are layered compounds with larger bandgaps than the MIR materials, enabling shift current response at higher energies. Their electronic band structures are shown in Figure~\ref{fig: band_IR}(d-h), while the corresponding shift current spectra in Figure~\ref{fig: sc_spectra_IR}(d-h) reveal strong shift current responses across a broad range from the NIR to the VIS spectrum. 

Overall, through high-throughput first principles calculations, we identified 9 IR shift current materials, seven of which were newly discovered. The strongest shift current response reaches 616 $\mu A/V^2$, which is four times greater than that of the previously reported IR-responsive topological semimetal TaAs~\cite{Osterhoudt2019_TaAs}. These discoveries establish a rich platform for IR detection, imaging, and energy conversion. Our work significantly expands the shift current material database into the IR spectral range, laying a theoretical foundation for subsequent experimental verification and device design.

\subsection*{Application of the dataset for AI-driven screening}
\begin{figure*}[!ht]
    \centering
    \includegraphics[width=0.98\linewidth]{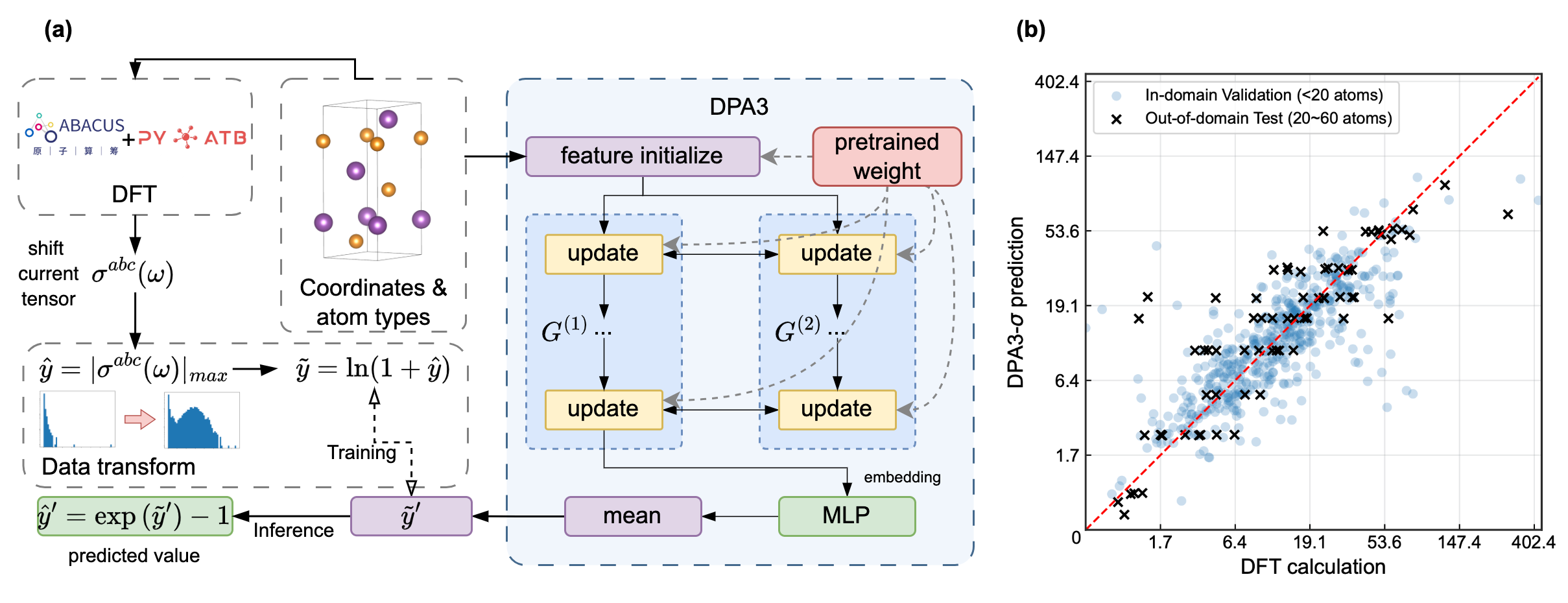}
    \caption{Architecture and performance of the DPA3-$\sigma$ model.(a) Schematic illustration of the training pipeline. The workflow integrates high-throughput DFT data generation with a transfer learning strategy, where the DPA3 backbone is initialized with pre-trained weights to extract geometric features. These features are mapped to the log-transformed maximum shift current via a Multi-Layer Perceptron (MLP). (b) Parity plot comparing DPA3-$\sigma$ predictions with DFT-calculated values. The plot distinguishes between the in-domain validation set (blue dots, $<20$ atoms) and the out-of-domain test set (black crosses, 20--60 atoms), demonstrating the model's generalizability to larger unseen systems. The red dashed line indicates perfect agreement. Due to the wide range of the data, the plot is constructed in the $\ln(1+\sigma)$ domain, while the axis tick labels are displayed in the original (real-value) scale.}
    \label{fig:dpa3}
\end{figure*}

The comprehensive dataset generated by our high-throughput screening serves as a foundational resource for data-driven discovery. As a demonstration, we utilized the DeepMD-kit~\cite{wang2018deepmd} to train our DFT database for the coarse screening of potential candidates with strong shift current response from the large-scale MP database, thereby narrowing the scope for subsequent precise calculations. 
Instead of training a model from scratch, we adopted a transfer learning strategy based on the DPA3 graph neural network (GNN)~\cite{wang2025graph}, a pre-trained large atomic model originally designed for interatomic potentials. 
DPA3 provides publicly available pre-trained weights and has demonstrated strong transferability as an atomistic encoder for materials. Thus, it is therefore adopted in this work to validate the feasibility of our pipeline for screening shift-current candidates.
We refer to this specialized shift-current predictor as \textbf{DPA3-$\sigma$}. Specifically, DPA3-$\sigma$ utilizes the pre-trained DPA3 parameters to generate high-quality structural embeddings, which are then mapped to the maximum shift current response $\sigma = |\sigma^{abc}(\omega)|_{max}$ via a multi-layer perceptron (MLP) (Figure~\ref{fig:dpa3}(a)). The DFT dataset (structures with $N_{atom} < 20$) was randomly split into training (80\%) and validation (20\%) sets. By leveraging the geometric features encoded in the pre-trained DPA3, DPA3-$\sigma$ significantly improves prediction efficiency on our finite dataset. We also applied a log-transformation ${y} = \ln(1+\sigma)$ to the target variable to mitigate the impact of large numerical variations.

\begin{table}[htbp!]
\centering
\setlength{\tabcolsep}{15pt}
\caption{Comprehensive performance evaluation of the DPA3-$\sigma$ model on the in-domain ($N=498$) and out-of-domain set ($N=73$). The upper table details the regression metrics in the logarithmic domain (${y} = \ln(1+\sigma)$), categorized by the magnitude of the DFT-calculated shift current. The lower table evaluates the model's utility as a binary screener using a threshold of 0.8 times the target values ($\sigma > 100~\mu A/V^2$ and $\sigma > 80~\mu A/V^2$). The high recall rate highlights the model's effectiveness in identifying top-tier candidates.}
\label{tab:ml_metrics}
\renewcommand{\arraystretch}{1.3}
\begin{tabular}{lcccc}
\toprule
\multicolumn{5}{c}{\textbf{Regression Metrics (Target: ${y} = \ln(1+\sigma)$)}} \\
\midrule
\textbf{Category} & \textbf{Small} & \textbf{Medium} & \textbf{Large} & \textbf{Overall} \\
Range ($\sigma$, $\mu A/V^2$) & $(0, 50]$ & $(50, 100]$ & $(100, \infty)$ & All \\
\midrule
Mean DFT (${y}$) & 2.43 & 4.14 & 5.44 & 2.56 \\
Mean Pred. (${y}'$) & 2.46 & 3.46 & 4.47 & 2.54 \\
MAE ($\downarrow$) & 0.34 & 0.72 & 0.97 & 0.37 \\
MAPE ($\downarrow$) (\%) & 19.33 & 17.30 & 17.00 & 19.18 \\
\midrule
\multicolumn{5}{c}{\textbf{Screening Performance (Threshold: Target $\times 0.8$)}} \\
\midrule
\textbf{Target} & \textbf{Recall} & \textbf{Precision} & \textbf{Accuracy} & \textbf{F1-Score} \\
$\sigma > 100~\mu A/V^2$ & \textbf{80.0\%} & 80.0\% & 99.6\% & 80.0\% \\
$\sigma > 80~\mu A/V^2$ & \textbf{85.7\%} & 60.0\% & 99.1\% & 70.6\% \\
\bottomrule
\end{tabular}
\end{table}

To rigorously assess the generalizability of DPA3-$\sigma$ beyond the training domain, we evaluated its performance on a combined dataset consisting of an \textbf{in-domain validation set} ($N=498$, $<20$ atoms) and an \textbf{out-of-domain test set} ($N=73$, $20-60$ atoms). As illustrated in Figure~\ref{fig:dpa3}(b), the model demonstrates high fidelity for in-domain structures (blue dots) and, crucially, maintains strong predictive correlation for larger, out-of-domain systems (black crosses). This confirms that the model has learned transferable geometric features rather than merely overfitting to small systems. Quantitative regression analysis (Table~\ref{tab:ml_metrics}, upper panel) further supports this observation. To assess the prediction accuracy in the logarithmic domain, we utilized the Mean Absolute Error (MAE) and Mean Absolute Percentage Error (MAPE), defined as:
$$
\text{MAE} = \frac{1}{n} \sum_{i=1}^{n} |{y}_i - {y}'_i|, \quad
\text{MAPE} = \frac{1}{n} \sum_{i=1}^{n} \left| \frac{{y}_i - {y}'_i}{{y}_i} \right| \times 100\%.
$$
where ${y}_i$ and ${y}'_i$ denote the DFT-calculated and predicted values, respectively. Using these metrics, we find that the model effectively differentiates response magnitudes across the diverse test set: the predicted mean for the "Large" group (4.47) is clearly distinct from the "Small" group (2.46). This significant separation ($\Delta \approx 2.0$), which far exceeds the overall MAE (0.37), validates the model's robust discriminative power for screening purposes.

To translate these predictive capabilities into a practical screening workflow, we evaluated the utility of DPA3-$\sigma$ as a binary classifier (Table~\ref{tab:ml_metrics}, lower panel). Given the model's tendency to slightly underestimate peak values (conservative prediction), we implemented a systematic "safety margin" strategy by setting the model's screening threshold to 0.8 times the desired target value (Threshold $= 0.8 \times$ Target). The factor of 0.8 was selected based on a sensitivity analysis with the target set to $100 ~ \mu A/V^2$, which provides the best balance between recall and precision. More details can be found in the Supplementary Materials  (Supplementary Table 3). We validated this strategy against two distinct targets. First, for our primary goal of identifying candidates with DFT-calculated responses $\sigma > 100~\mu A/V^2$, we set the model threshold to $80~\mu A/V^2$. This approach yielded an impressive Recall of 80.0\%, meaning it successfully retrieved 4 out of 5 true high-performance candidates, and a Precision of 80.0\%, indicating a high "hit rate" that minimizes wasted calculations on false positives. Second, to assess robustness on a broader set of candidates, we targeted materials with responses $\sigma > 80~\mu A/V^2$, which achieved an even higher Recall of 85.7\%. These consistent results confirm that the DPA3-$\sigma$ can reliably retrieve the vast majority of high-performance materials while maintaining high efficiency.

Finally, to demonstrate the practical value of this AI-driven workflow, we applied DPA3-$\sigma$ to screen the remaining 6,008 non-centrosymmetric materials in the MP database (Supplementary Figure 6). We specifically targeted complex systems with more than 30 atoms—a regime computationally expensive for exhaustive DFT screening. From the top predictions, we selected 10 candidates for validation, successfully identifying two new large-cell materials, Cs(Bi$_2$Te$_3$)$_2$ (mp-672338) and Sn(BiTe$2$)$_2$ (mp-677596), with confirmable high responses (Supplementary Table 2). Remarkably, this successful discovery using a simplified scalar target ($|\sigma^{abc}|_{max}$) serves as a compelling proof-of-concept, demonstrating that our high-quality dataset enables effective AI-driven material design even with straightforward modeling strategies.

\section*{Discussion}

The shift current is an intrinsic second-order photocurrent arising from the quantum geometry of electronic states in non-centrosymmetric materials. As reflected in Eq. (7) of the methods section, its magnitude is governed by the interplay between interband optical coupling and the geometric properties of Bloch wavefunctions~\cite{PhysRevB.61.5337,PhysRevLett.109.116601}. Our high-throughput screening reveals that large shift current responses can be attributed to the synergistic combination of $C_{3v}$ symmetry and heavy $p$-block chemistry. Unlike $d$-orbitals, which are typically localized, or $s$-orbitals, which lack intrinsic directionality, the heavy $p$-orbitals (e.g., Te-5$p$, Sb-5$p$) possess both strong directionality and significant spatial delocalization. This allows for the interband transition over a longer spatial distance, leading to a large shift vector magnitude. Besides, their extended $p$ orbitals dominate the band edges, yielding large optical transition matrix elements due to strong orbital overlap. In parallel, polar $C_{3v}$ symmetry provides a preferred current direction and nontrivial Berry connections, and symmetry constraints concentrate the nonlinear response into a small number of tensor components, thereby reducing cancellation effects. Many high-performing $C_{3v}$ materials further adopt layered trigonal structures, where pronounced structural anisotropy enhances the asymmetry of photoexcited states, jointly leading to enhanced shift current responses.

It should be clarified that all calculations in this work are performed without spin-orbit coupling (SOC). In systems containing heavy elements, SOC is expected to further modify the band structure and Berry-phase related quantities, potentially shifting resonant energies and modulating the shift current response~\cite{2016_sciadv.1501524}. However, given that the primary objective of this work is high-throughput screening rather than quantitative refinement of individual materials, SOC effects are not included here and are instead identified as an important direction for future, more detailed investigations. Overall, the synergistic combination of polar $C_{3v}$ symmetry and heavy $p$-block chemistry optimizes the key ingredients governing shift current generation, providing a clear microscopic rationale for the large responses identified here and offering practical guidance for discovering high-performance infrared shift current materials.

Assessing experimental feasibility is as critical as identifying materials with strong shift current responses, and we thus adopted the universal thermodynamic stability criterion of $E_{hull}$ < 50 meV/atom in a high-throughput framework to evaluate synthesizability  of candidate materials. Notably, several top-performing candidates feature well-established experimental synthesis routes that confirm their practical accessibility, with Ge–Sb–Te system compounds such as GeSb$_2$Te$_4$ growable as crystalline films or bulk single crystals via vapor transport and related methods, and widely studied for their phase-change and optoelectronic properties in chalcogenide research~\cite{ZHEZHU2025109863}. Binary chalcogenides (e.g., GeTe) and layered tellurides (e.g., Sb$_2$Te$_3$) are also well-documented, synthesizable by conventional solid-state or hydrothermal methods with abundant reports on growing high-quality bulk crystals and nanostructures~\cite{C5RA20014H}, and these systems further form the foundation of Ge–Sb–Te alloys, whose synthesis and characterization have been extensively elaborated~\cite{Matsunaga:og0002}. Moreover, the commercial availability of single crystals for several related chalcogenides, such as GeSb$_2$Te$_4$, GeTe, PbBi$_2$Te$_4$, and Sb$_2$Te$_2$Se, provides direct evidence of experimental feasibility for many of the high-performance materials identified here. The close chemical relationship between our top candidates and widely studied chalcogenide materials suggests that many of the predicted high shift-current materials are experimentally accessible, even in the absence of specific synthesis reports. Together with our thermodynamic stability filtering, this indicates the screened chemical space as chemically realistic and experimentally viable, thus motivating future experimental validation and device exploration of these materials.

In this work, we screened 2,519 candidate materials with broken inversion symmetry by applying a multi-step filter to the the MP database, systematically evaluating their shift current responses using high-throughput DFT calculations. Ultimately, 32 NLO materials with significant shift current responses ($\sigma$ > 100 $\mu A/V^2$) were identified, among which 9 exhibit activity in the IR region. Notably, Sn$_5$Ge$_2$(SbTe$_5$)$_2$ (mp-1219067) stands out as a top performer with an exceptional shift current response of 616 $\mu A/V^2$, which is four times greater than that of the previously reported topological semimetal TaAs~\cite{Osterhoudt2019_TaAs}. The majority of these materials possess layered structures with $C_{3v}$ symmetry and heavy $p$-block elements (e.g. Te, Sb), featuring narrow bandgaps, strong NLO response, and favorable thermodynamic stability. These structural and electronic attributes endow them with broad application potential in IR photodetection, bias-free self-driven sensing, broadband energy harvesting, and deep biomedical modulation. 

A key advance of this study is that it addresses a critical gap in NLO material screening by successfully extending the shift current operational window from the VIS/UV into the IR spectral region. 
Beyond significantly enhancing the efficiency and scope of materials discovery, the high-throughput DFT approach employed here establishes comprehensive, high-quality datasets that underpin data-driven material design and machine-learning research. As a proof of concept, we validated the feasibility of the DPA3-$\sigma$-based methods for rapid preliminary screening of material databases. 

Collectively, this work establishes a scalable, multi-stage screening framework that integrates high-throughput first-principles calculations and machine-learning-assisted prediction, which can be readily extended to other materials databases or NLO responses. Future research can proceed along two key directions. First, conduct SOC-inclusive first-principles calculations or targeted experimental verification on the high-performance candidates identified in this work. Second, leveraging the comprehensive shift current tensor dataset established in this study, develop generative models to discover additional high-performance materials or construct predictive models capable of predicting the full, frequency-dependent shift current tensor. These follow-up studies will further refine and verify the proposed screening framework, accelerating the translation of these computationally predicted NLO materials into practical optoelectronic devices.

\section*{Methods}

\subsection*{Theory of the shift current response}
The shift current constitutes an essential mechanism underlying the BPVE in non-centrosymmetric materials~\cite{PhysRevB.61.5337}. It describes the photocurrent generated by light illumination on homogeneous crystals without inversion symmetry. As a second-order NLO response, the shift current can be expressed as a direct current induced by a monochromatic light excitation with the form of $\mathbf{E}(t)=\mathbf{E}(\omega) \mathrm{e}^{i \omega t}+\mathbf{E}(-\omega) \mathrm{e}^{-i \omega t}$, where
\begin{equation}
    J_{shift}^{a}=2\sigma^{a b c}(0;\omega, -\omega) E^{b}(\omega) E^{c}(-\omega)
\end{equation}
where $a$, $b$, $c$ are Cartesian indices, with $a$ specifying the current direction and $b$ and $c$ incident light  polarization.
The shift current conductivity $\sigma^{abc}$ is given by 
\begin{equation}
    \begin{aligned}
\sigma^{a b c}(0 ; \omega,-\omega)= & -\frac{i \pi e^{3}}{2 \hbar^{2}} \int \frac{d \boldsymbol{k}}{(2\pi)^3} \sum_{n, m} f_{n m}\left(I_{m n}^{a b c}+I_{m n}^{a c b}\right) \times \delta\left(\omega_{m n}-\omega\right)
\end{aligned}
\label{shift-current tensor}
\end{equation}
where $\hbar \omega _{nm} = E_n -E_m$ represents photon energy, $f_{nm} = f(E_{n}) - f(E_{m})$ is the Fermi-Dirac occupation number. The integrand $I_{n m}^{a b c}$ can be written out with Berry connections and derivatives of Berry connections
\begin{equation}
    I_{n m}^{a b c} = r_{m n}^{b}r_{n m; a}^{c}
\end{equation}
where $r_{n m}^{b}$ and $r_{n m; a}^{b}$ are given by 
\begin{equation}
    r_{n m}^{b} = (1 - \delta_{n m})A_{n m}^{b}
\end{equation}
and
\begin{equation}
    r_{n m; a}^{b} = \partial_{k_a} r_{n m}^{b}-i\left(A_{n n}^{a}-A_{m m}^{a}\right) r_{n m}^{b}
\end{equation}
respectively, here the Berry connection is defined as
\begin{equation}
    A_{n m}^{b} \equiv i \left\langle u_{n} \mid \partial_{b} u_{m}\right\rangle
\end{equation}
where $\left | u_m \right \rangle$ denotes the periodic part of a Block eigenstate. Since a $\delta$ function is present in the expression of $\sigma^{a b c}$, a very fine $\boldsymbol{k}$ mesh is needed to achieve convergence.

Under linearly polarized light ($b = c$), equation (\ref{shift-current tensor}) can be reformulated into a more transparent form 

\begin{equation}
    \begin{aligned}
\sigma^{a b b}(0 ; \omega,-\omega)= & \frac{\pi e^{3}}{\hbar^{2}} \int[d \boldsymbol{k}] \sum_{n, m} f_{n m} R_{n m}^{a,b} \times r_{n m}^{b} r_{m n}^{b} \times \delta\left(\omega_{m n}-\omega\right),
\end{aligned}
\end{equation}
where $R_{n m}^{a,b}$ is shift vector defined as

\begin{equation}
    R_{n m}^{a,b}=\partial_{k_a} \phi_{n m}^{b}-A_{m m}^{a}+A_{n n}^{a}
\end{equation}
and $\phi _{n m}^{b}$ is the phase of $r_{n m}^{b}=\left|r_{n m}^{b}\right| e^{-i \phi_{n m}^{b}}$. $R_{n m}^{a,b}$ has a unit of length and can be physically interpreted as the average displacement of the coherent photoexcited carriers during their lifetimes. The product $r_{n m}^{b} r_{m n}^{b} \delta\left(\omega_{n m}-\omega\right)$
can be interpreted as the transition rate from band $m$ to band $n$ according to the Fermi golden rule. That is, shift current tensor can be expressed as transition rate from band $m$ to band $n$ multiplied by shift vector between the two bands.

\subsection*{Definition of large shift current}

The maximum value of the shift current tensor element $|\sigma^{abc}| _{max}$ was adopted as the performance metric to define large shift current, as this scalar reflects the intrinsic upper limit of device performance in technologically relevant scenarios where light polarization and crystallographic orientation can be controlled to align with the tensor’s strongest component, and it avoids diluting the strong anisotropy of shift current responses—an issue that would arise with averaged tensor quantities and risk excluding promising candidate materials during screening.
In this work, we set $|\sigma^{abc}| _{max} \ge 100 ~\mu A/V^2$ as the threshold to identify materials with strong shift current response (note that $|\sigma^{abc}| _{max}$ is sometimes used interchangeably with $\sigma$ in the text). This criterion is motivated by established benchmarks in the literature. Conventional bulk photovoltaic materials, such as ferroelectric oxides BaTiO$_3$ ($\sigma \sim 30 ~\mu A/V^2$) and PbTiO$_3$ ($\sigma \sim 50 ~\mu A/V^2$), as well as the non-centrosymmetric semiconductor GaAs ($\sigma \sim 40 ~\mu A/V^2$), typically exhibit relatively modest shift current responses. In contrast, enhanced responses on the order of $\sim 100 ~\mu A/V^2$ have been reported in low-dimensional systems such as monolayer GeS and WS$_2$, while even larger values ($\sigma \sim 150-200 ~\mu A/V^2$) are observed in topological semimetals such as TaAs. Therefore, choosing 100 $\mu A/V^2$ as the screening threshold effectively distinguishes materials with shift current responses that clearly exceed those of traditional bulk photovoltaic compounds, while remaining consistent with the magnitude achieved in known high-performance systems.
\\

\subsection*{DFT calculation details}

We started with the structures provided by the MP database and performed first-principles DFT calculations based on the numerical atomic bases (NAOs), which is more suitable for calculations of large systems~\cite{LI2016503}. All DFT calculations were carried out using the ABACUS package~\cite{Chen_2010, Zhou2025_abacus}. The charge density was converged to within $1 \times 10^{-8}$ $e$/a.u.$^3$ and a uniform $k$-point mesh with a spacing of 0.03 $\times$ 2$\pi$/Å for self-consistent field (SCF) calculations. A Gaussian smearing scheme with a broadening of 0.01 Ry was adopted throughout this work to ensure numerical stability in Brillouin-zone integrations. In ABACUS, we employed double-zeta polarized (DZP) pseudo-atomic orbital basis set to balance computational accuracy and efficiency. The SG15 optimized Norm-Conserving Vanderbilt (ONCV) multi-projector pseudopotentials~\cite{sg15_oncv_2015} and the atomic basis~\cite{NAO_2021PRB} paired with the SG15 pseudopotential are used.
We verified that our basis sets for high-throughput calculations yield similar band structures to those provided by the MP database. 
For exchange-correlation effects, the PBE functional~\cite{1996_PBE} was adopted for the filtered 2,519 materials. For the top 32 candidates, we further employed the Heyd-Scuseria-Ernzerhof 2006 (HSE06) screened hybrid functional~\cite{2003_HSE, Lin2020_hse}, which generally provides more accurate description of electronic band gaps. For the out-of-domain test set with more than 20 atoms per unit cell, all first-principles evaluation were performed at the PBE functional.

Optical responses, such as the second-order shift current tensor $\sigma$, were calculated using the Python Ab-initio Tight-Binding (PYATB) package~\cite{JIN2023108844} based on the DFT Hamiltonian obtained from ABACUS. We found that a $k$-point mesh spacing of 0.02 $\times$ 2$\pi$/Å is sufficient for the convergence of frequency-dependent shift current spectra for the majority of materials, while a denser $k$-mesh is necessary for few cases (e.g. HfGeTe$_4$, mp-567817). The convergence of Gaussian smearing and $k$-point sampling has been systematically tested, and the corresponding results are provided in Supplementary Figures 4 and 5. For selected top candidate materials, the shift current response was recalculated using the full HSE electronic structure, including both band energies and wavefunctions obtained self-consistently within the HSE functional. No scissor correction was applied to PBE results. This approach ensures a consistent and accurate description of the optical transition matrix elements and Berry-phase related quantities that govern the shift current response. 

\section*{Acknowledgements}
This work was supported by the National Natural Science Foundation of China (Grant No. 12125504 and No. 12504285), the National Key R\&D Program of China (2022YFA1404400), and the ``Hundred Talents Program'' of the Chinese Academy of Sciences, and the Priority Academic Program Development (PAPD) of Jiangsu Higher Education Institutions, and the Natural Science Foundation of Jiangsu Province (BK20250472). We are grateful for the insightful discussion with Dr. Gan Jin regarding the application of the PYATB software.

\bibliography{refs}

\section*{Data availability}
The DPA3-$\sigma$ model weights are publicly available at \url{https://huggingface.co/CLaS/DPA3-shift_current}.
The data supporting the findings of this study are available within the article and its supplementary materials.

\section*{Competing interest}
The authors declare no competing interests.

\section*{Author contributions statement}

J.-H. Jiang initiated the study. J.-H. Jiang, Q. Gu, Q. Wang and D. Zheng co-supervised the project. A. Yang performed the high-throughput frist-principles calculations. D. Jin and M. Liu performed the machine-learning demonstration.
A. Yang and D. Jin analyzed and visualized the results. A. Yang, D. Jin, and Q. Gu wrote the manuscript. All authors reviewed the manuscript.

\end{document}